# Resolution of hyperfine transitions in metastable $^{83}$Kr using Electromagnetically Induced Transparency

Y. B. Kale*, S. R. Mishra, V. B. Tiwari, S. Singh and H. S. Rawat

Laser Physics Applications Section,
Raja Rammanna Centre for Advanced Technology, Indore, - 452013, India.

*e-mail: yogeshwar@rrcat.gov.in

**Abstract**

Narrow linewidth signals of Electromagnetically Induced Transparency (EIT) in the metastable $^{83}$Kr have been observed for the first time. Various hyperfine transitions in $4p^55s[3/2]_2$ to $4p^55p[5/2]_3$ manifolds of $^{83}$Kr have been identified through the experimentally observed EIT signals. Some unresolved or poorly resolved hyperfine transitions in saturated absorption spectroscopy (SAS) are clearly resolved in the present work. Using the spectral separation of these EIT identified hyperfine transitions, the magnetic hyperfine constant (A) and the electric quadrupole hyperfine constant (B) are determined with improved accuracy for $4p^55s[3/2]_2$ and $4p^55p[5/2]_3$ manifolds.

# I. INTRODUCTION

The phenomenon of electromagnetically induced transparency (EIT) was introduced in the last century [1]. The EIT represents the reduction in absorption of light by a medium due to quantum interference among different transition amplitudes in a multi-state system. The quantum interference among these transition amplitudes may result in net reduction in the absorption between two states if one of these two states is coupled to the other states. This results in the formation of a transparency window in the absorption profile corresponding to a resonant excitation [2]. The EIT phenomenon has been demonstrated in several systems which include Doppler broadened gaseous media [3, 4], cold atoms samples [5, 6], nonlinear crystals [7], metamaterials [8], cold Rydberg atoms [9] etc. The EIT phenomenon has also been observed in metastable noble gas atoms [10, 11] and in optomechanical resonators [12]. The same is predicted theoretically where authors have considered an atomic medium that is initially prepared in a spin-wave or superatom state [13]. In Ref. [13], the EIT with two relevant Rydberg states is investigated where excitations can be exchanged between distant atoms. Recently, EIT has also been used for laser cooling of the motional modes of ion chain [14]. Ultra-narrow EIT signals have been obtained in paraffin coated vapor cells by reducing the ground state decoherences [15]. The narrow EIT spectra have expanded their horizon in various fields like slow light propagation [16], quantum storage [17], ultrasensitive magnetometery [18] and atomic frequency offset locking [19]. The atomic coherences generated by EIT also play crucial role in nonlinear optics [20] and quantum memory [21]. The use of EIT to enhance four wave mixing in Krypton (Kr) atoms has been discussed earlier [22].

Here we report the narrow EIT peaks observed in the transmitted probe beam signal in pump-probe spectroscopy of metastable $^{83}$Kr ($^{83}$Kr*) atoms for the first time. These narrow EIT signals have been used to identify various hyperfine transitions in $4p^55s[3/2]_2$ to $4p^55p[5/2]_3$ manifolds of $^{83}$Kr* atom. Some transitions which are unresolved or poorly resolved in the well known saturated absorption spectroscopy (SAS) are clearly resolved in the present work based on the EIT technique. Since $^{83}$Kr is the only high abundance fermionic isotope, it is preferred to use $^{83}$Kr* as a frequency reference for the preparation of laser cooled samples of $^{85}$Kr* and $^{81}$Kr* which find their applications in fields like Atom Trap Trace Analysis (ATTA) [23]. Thus, it is important to resolve the transitions of $^{83}$Kr* with the frequency uncertainty ~1 MHz or less. It is already identified that the closed transitions of $^{85}$Kr* and $^{81}$Kr* for cooling purpose are separated by ~ 87 MHz and ~ 21 MHz respectively from 13/2-15/2 and 11/2-11/2 transitions of $^{83}$Kr* atom in $4p^55s[3/2]_2$ to $4p^55p[5/2]_3$ manifolds [24].

## II. RESOLUTION OF TRANSITIONS USING EIT:

In the natural Krypton (Kr) gas sample which we used, all the isotopes are present according to their natural abundances (e.g. $^{84}$Kr (56.9%), $^{86}$Kr (17.3%), $^{82}$Kr (11.6%), $^{83}$Kr (11.5%)) [25]. In the SAS of metastable Kr atoms, the Doppler broadened absorption profiles of all these isotopes overlap with one another. The Doppler absorption profiles of $^{84}$Kr and $^{86}$Kr metastable atoms are ~24.5 and ~2.3 times respectively larger in magnitude than that of metastable $^{83}$Kr. Thus, spectral resolution in SAS of metastable $^{83}$Kr becomes poor when - (i) The transitions are buried under the Doppler broadened absorption profiles/ Doppler pedestals of highly abundant isotopes, (ii) the transitions are closely spaced in spectrum and (iii) the transitions are weak.

We have used EIT in three-level system as an alternative technique to resolve the hyperfine transitions in metastable $^{83}$Kr atom. The EIT phenomenon is based on quantum interference effect and can provide subnatural linewidth that is sensitive to control beam parameters, it is expected to obtain higher resolution and better sensitivity using the EIT than the population based SAS technique. The following discussion explains our approach to use EIT for the resolution of hyperfine transitions in metastable $^{83}$Kr atom.

We consider a three-level Λ-type or V-type atomic system (see Fig. 1) for EIT purpose in which there are two dipole allowed transitions ($|1\rangle$ to $|2\rangle$ and $|3\rangle$ to $|2\rangle$) and one dipole forbidden transition ($|1\rangle$ to $|3\rangle$). In these systems, we assume that only one transition (say $|1\rangle$ to $|2\rangle$) out of two dipole allowed transitions is resolvable by SAS. We then perform EIT measurements using probe and control laser beams in copropagating geometry in the above three level system. Here the probe laser frequency is kept at the peak (i.e. probe laser detuning $\Delta_P = 0$) of the resolved transition (using a SAS signal for reference) and the control laser frequency is scanned around the unresolved transition to record the variation in transmitted probe signal. The probe transmission will give the EIT peak when the control laser frequency is equal to the resonance frequency of unresolved transition. This is because EIT can be observed when three level (Λ or V type) system satisfies the two-photon resonance condition (i.e. $\Delta_P = \Delta_C$, where, $\Delta_C$ is the control laser detuning from unresolved transition) [2]. The linewidth of Λ system EIT in Doppler broadened medium, can be written as $\Gamma_{EIT} = [2\Gamma_{13}/\Gamma]^{1/2} \Omega_C$, which can be smaller for the smaller control laser Rabi frequencies $\Omega_C$ and relaxation rate satisfying the condition $\Gamma_{13} \ll \Gamma$ [26]. The linewidth of SAS resonance is given by $\Gamma_{SAS} = \Gamma[1+(\Omega_C^2/\Omega_S^2)]^{1/2}$ which is always limited by natural linewidth ($\Gamma$), where $\Omega_S$ is the Rabi frequency corresponds to saturation intensity of the transition. Hence, EIT can give peaks of narrower linewidth than those obtained in SAS [27]. The amplitude of EIT signal can also be controlled by changing the intensity of the control laser. Thus, EIT can be used in precise and accurate resolution of the unresolved resonances.

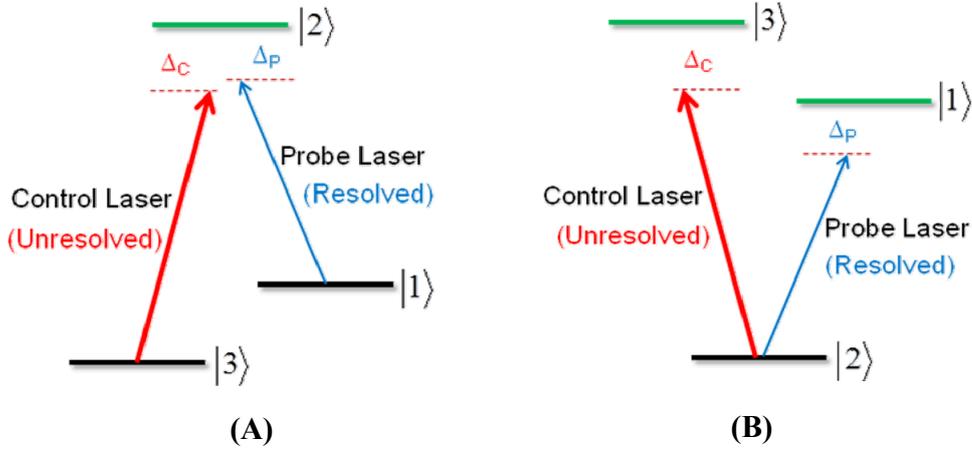

*Fig. 1: Schematics of three level systems used for EIT purpose: (A) Λ-type and (B) V-type. Here transitions $|1\rangle$ to $|2\rangle$ and $|3\rangle$ to $|2\rangle$ are dipole allowed and $|1\rangle$ to $|3\rangle$ is dipole forbidden transition. $\Delta_P$ ($\Delta_C$) is probe (control) laser frequency detuning.*

The $^{83}$Kr has nuclear spin I = 9/2. Therefore, for the $4p^55s[3/2]_2$ manifold having angular momentum J = 2, there are five hyperfine levels with F values varying from 5/2 to 13/2. Similarly, for $4p^55p[5/2]_3$ manifold having J' = 3, the number of hyperfine levels are seven with F' varying from 3/2 to 15/2. Thus, there are 15 possible principle transitions and 15 crossover transitions which result in the 'forest' of peaks in SAS spectrum. Out of these 15 principle transitions, only seven transitions are clearly resolved in SAS signal whereas rest of the transitions are either poorly resolved or unresolved in the SAS of $^{83}$Kr. The experimentally obtained SAS signals are shown in Fig. 2 (A) and the corresponding hyperfine transitions in $4p^55s[3/2]_2$ to $4p^55p[5/2]_3$ manifolds in $^{83}$Kr* are shown in Fig. 2 (B). Some transitions among these, shown by dotted lines (denoted as F-F') such as 7/2 - 5/2, 9/2 - 7/2, 9/2 - 9/2 and 11/2 -

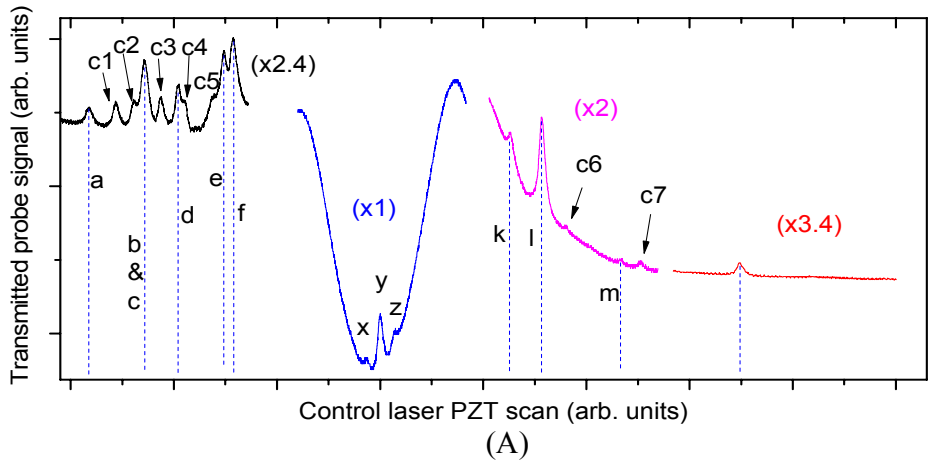

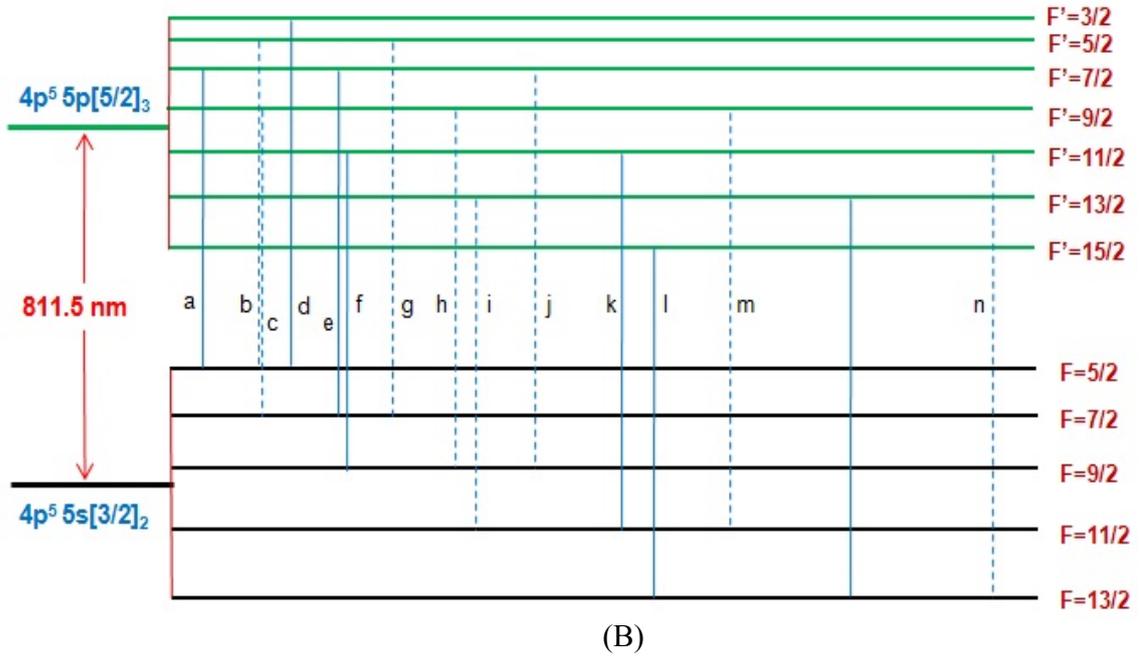

*Fig.2: (A) The observed SAS signals from SAS-c setup. (B) Various hyperfine transitions in $4p^55s[3/2]_2$ to $4p^55p[5/2]_3$ manifolds of $^{83}Kr^*$. In (A), a to n are the transition peaks and $C_n$ are the crossover peaks of $^{83}Kr^*$ hyperfine transitions. The peaks x, y and z are SAS peaks of $^{82}Kr^*$, $^{84}Kr^*$ and $^{86}Kr^*$ isotopes respectively in $4p^55s[3/2]_2$ to $4p^55p[5/2]_3$. The y-axis magnifications used in plotting the signal are indicated in parentheses. The dotted lines in (B) indicate the transitions which are not clearly resolved in SAS-c spectrum shown in (A).*

13/2 in Fig. 2(B), are not resolved because these transitions are buried under the large Doppler backgrounds/pedestals of highly abundant isotopes $^{84}Kr$ and $^{86}Kr$. Some other transitions such as, 5/2 - 5/2 and 7/2 - 9/2 in $^{83}Kr^*$ that are closely separated, remain unresolved in SAS spectrum. Finally, there are also some transitions which are either weak and/or open due to small transition strength (S) and/or branching ratio (b). Such transitions, for example 11/2 - 9/2 with S = 0.06 and

b = 0.1 and 13/2 - 11/2 with S = 0.02 and b = 0.03, are also not resolved in SAS spectrum. The values of S and b for each of the transition in $4p^55s[3/2]_2$ to $4p^55p[5/2]_3$ manifolds are given by following equations [28, 29],

$$S_{F' \to F} = (2F'+1)(2J+1)\begin{Bmatrix} J & J' & 1 \\ F' & F & I \end{Bmatrix}^2 \quad (1)$$

$$b_{F \to F'} = (2F+1)(2J'+1)\begin{Bmatrix} F' & 1 & F \\ J & I & J' \end{Bmatrix}^2 \quad (2)$$

*Table I: Transition strengths and branching ratios for the hyperfine transitions in $4p^55s[3/2]_2$ to $4p^55p[5/2]_3$ manifolds of $^{83}$Kr.*

| No. | Transition F - F' | Transition Strength (S) | Transition Branching Ratio (b) |
|---|---|---|---|
| 1 | 13/2 - 15/2 | 0.82 | 1 (closed) |
| 2 | 13/2 - 13/2 | 0.16 | 0.23 |
| 3 | 13/2 - 11/2 | 0.02 | 0.03 |
| 4 | 11/2 - 13/2 | 0.64 | 0.77 |
| 5 | 11/2 - 11/2 | 0.30 | 0.42 |
| 6 | 11/2 - 9/2 | 0.06 | 0.1 |
| 7 | 9/2 - 11/2 | 0.47 | 0.55 |
| 8 | 9/2 - 9/2 | 0.39 | 0.55 |
| 9 | 9/2 - 7/2 | 0.13 | 0.23 |
| 10 | 7/2 - 9/2 | 0.31 | 0.35 |
| 11 | 7/2 - 7/2 | 0.44 | 0.61 |
| 12 | 7/2 - 5/2 | 0.26 | 0.48 |
| 13 | 5/2 - 7/2 | 0.15 | 0.16 |
| 14 | 5/2 - 5/2 | 0.37 | 0.52 |
| 15 | 5/2 - 3/2 | 0.48 | 1 (closed) |

where, the term in curly bracket in right hand side of Eq. (1) and (2) is the square of the 6-j symbol. The calculated values of S and b for each of the transition in $4p^55s[3/2]_2$ to $4p^55p[5/2]_3$ manifolds are listed in Table I.

We demonstrate in this work that the unresolved transitions as discussed before can be resolved using EIT in three level systems (mainly Λ and V). From the measured frequency positions of experimentally observed EIT signals, the transition frequencies of various hyperfine transitions from lower ($4p^55s[3/2]_2$) to upper ($4p^55p[5/2]_3$) manifolds in $^{83}$Kr* are obtained. Using this data, the hyperfine splitting for both of these manifolds have been estimated, and corresponding magnetic hyperfine constant (A) and electric quadrupole hyperfine constant (B) for both the manifolds are evaluated.

### III. EXPERIMENTAL SETUP

Fig. 3 shows the schematic of the experimental setup. Two external cavity diode laser (ECDL) systems (Toptica, Germany) with wavelength ~ 811.5 nm and linewidth ~ 700 kHz have been used as control and probe lasers to perform EIT experiments. The SAS-c and SAS-p are the SAS setups for the frequency referencing of control and probe laser beams. The control (strong) and the probe (weak) laser beams ($1/e^2$ spot size ~ 0.3 mm) are in co-propagating geometry and are carefully merged (separated) using the polarizing beam splitters PBS1 and PBS2. Slight misalignment is introduced in control and probe beams to restrict the residual intensity of strong control beam falling on the photodetector (PD1). This is also helpful to avoid beating between the probe laser beam and the residual control laser beam falling on the detector [30]. The Kr gas cell (pressure ~ 200 mTorr) used for EIT experiments is placed in between the pair of the PBSs. The gas cell is kept inside a multi-turn copper coil used for RF excitation (with frequency ~ 30 MHz) of Kr gas. The coil and cell assembly is surrounded by a cylindrical μ-metal shield to avoid the influence of stray magnetic fields. The combinations of PBS and half waveplate (λ/2) are used to control the intensities of the control laser (Rabi frequency $\Omega_C$ is varied from 4.5 to 8.0 MHz) and probe laser (Rabi frequency $\Omega_P$ is fixed at ~ 0.4 MHz throughout the experiment) beams passing through the Kr gas cell. We note here that the Kr gas cells and the RF coils used in both the SAS setups are identical to that used for recording the EIT signals. In our experiments, the EIT signal is obtained by measuring the variation in the transmitted probe beam signal when the probe laser frequency is fixed at a resolved transition (known from the SAS-p spectrum) and the control laser frequency is scanned around the transition to be investigated.

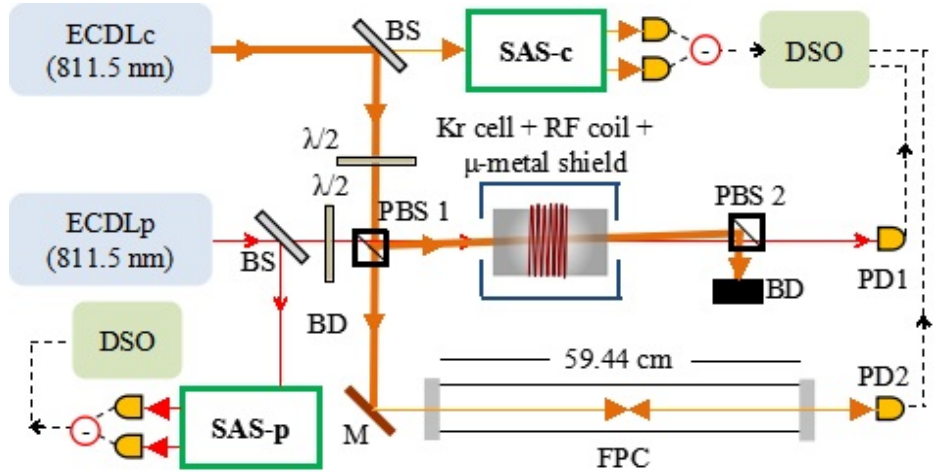

*Fig. 3: The schematics of the experimental setup. ECDLc(p): control(probe) laser, SAS-c(-p): saturated absorption spectroscopic setups for control (probe) laser, λ/2: half waveplate, PBS: polarizing beam splitter, M: mirror, FPC: Fabry-Pérot Cavity, DSO: digital storage oscilloscope, PDs: photodetectors, BS: Beam splitter, BD: beam dump.*

It is known that a linear voltage ramp applied to the piezo-electric transducer (PZT) attached to the cavity of an ECDL results in non-linear variation of laser frequency with the voltage at higher values of applied voltage [31]. Thus PZT voltage scan is not suitable to directly read the change in laser frequency. To overcome this difficulty, a part of control laser beam was passed through a passive Fabry-Pérot Cavity (FPC) and its transmission peaks were recorded during the PZT scan of the control laser. After knowing the free spectral range (FSR) of FPC, and counting the number of transmission peaks (teeth of comb), any duration length in the PZT scan could be converted into frequency range. By taking the frequency separation between $^{84}$Kr* transition peak and $^{83}$Kr* closed transition (13/2 - 15/2) peak equal to 783 MHz in the SAS spectrum [24], the data in Fig. 4 gives the average FSR of FPC to be ~252.2 MHz over four teeth in that spectral range. Fig. 4 illustrates the method of calibration of the control laser PZT scan. The resolution of our recorded spectra is ~0.9 MHz which is the separation of two consecutive data points in the oscilloscope trace. The trace (a) shows the FPC output, trace (b) shows the SAS signal of control laser, trace (c) shows the recorded EIT spectra and trace (d) shows the the PZT voltage (on 1/10 scale). All the traces are recorded simultaneously by applying a linear voltage ramp to the PZT of the control laser. In Fig. 4, the probe laser is kept at 9/2 - 11/2 transition using the SAS-p. This results in the formation of narrow Λ-type EITs in probe absorption at the frequency positions corresponding to 11/2 - 11/2 and 13/2 - 11/2 transitions (trace (c) in Fig. 4). The other EIT spectra in our experiments have been recorded in similar manner by keeping probe laser frequency at the resolved transitions using SAS-p and scanning the control laser frequency. The frequency positions of various EIT peaks are obtained from the calibration of the control laser PZT scan.

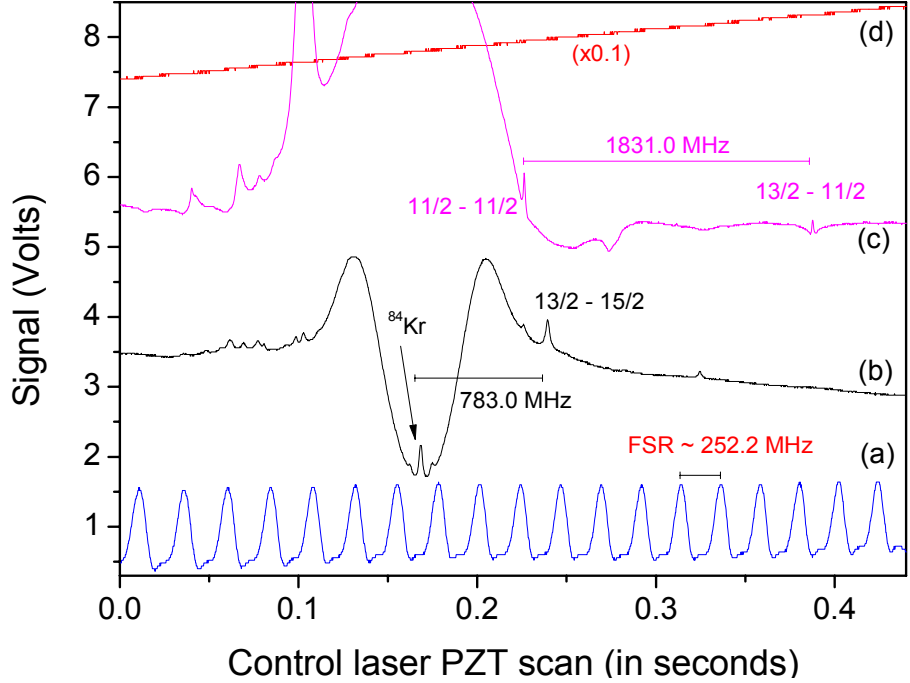

*Fig. 4: Simultaneously recorded traces using digital storage oscilloscope (DSO). (a) Fabry-Pérot Cavity output, (b) control laser SAS signal, (c) EIT signal and (d) PZT voltage.*

## IV. RESULTS AND DISCUSSION

Fig. 5 (A) shows the various hyperfine transitions in $4p^55s[3/2]_2$ to $4p^55p[5/2]_3$ manifolds of metastable $^{83}$Kr atom and Fig. 5 (B) shows the observed EIT signals corresponding to these hyperfine transitions. The hyperfine transitions indicated by red vertical lines in Fig. 5 (A) could not be observed in SAS spectrum of the control laser (Fig. 2 (A)). It is evident from Fig. 5 (B) that these unresolved transitions in SAS are clearly resolved in the EIT spectra (refer peaks b, c, g, h, i, j, m and n in Fig. 5 (B)). The peaks a and c' are Λ- and V-type EIT resonances obtained at frequency positions corresponding to 5/2 - 7/2 and 7/2 - 9/2 transitions respectively. These signals are obtained after scanning the control laser frequency and recording the probe

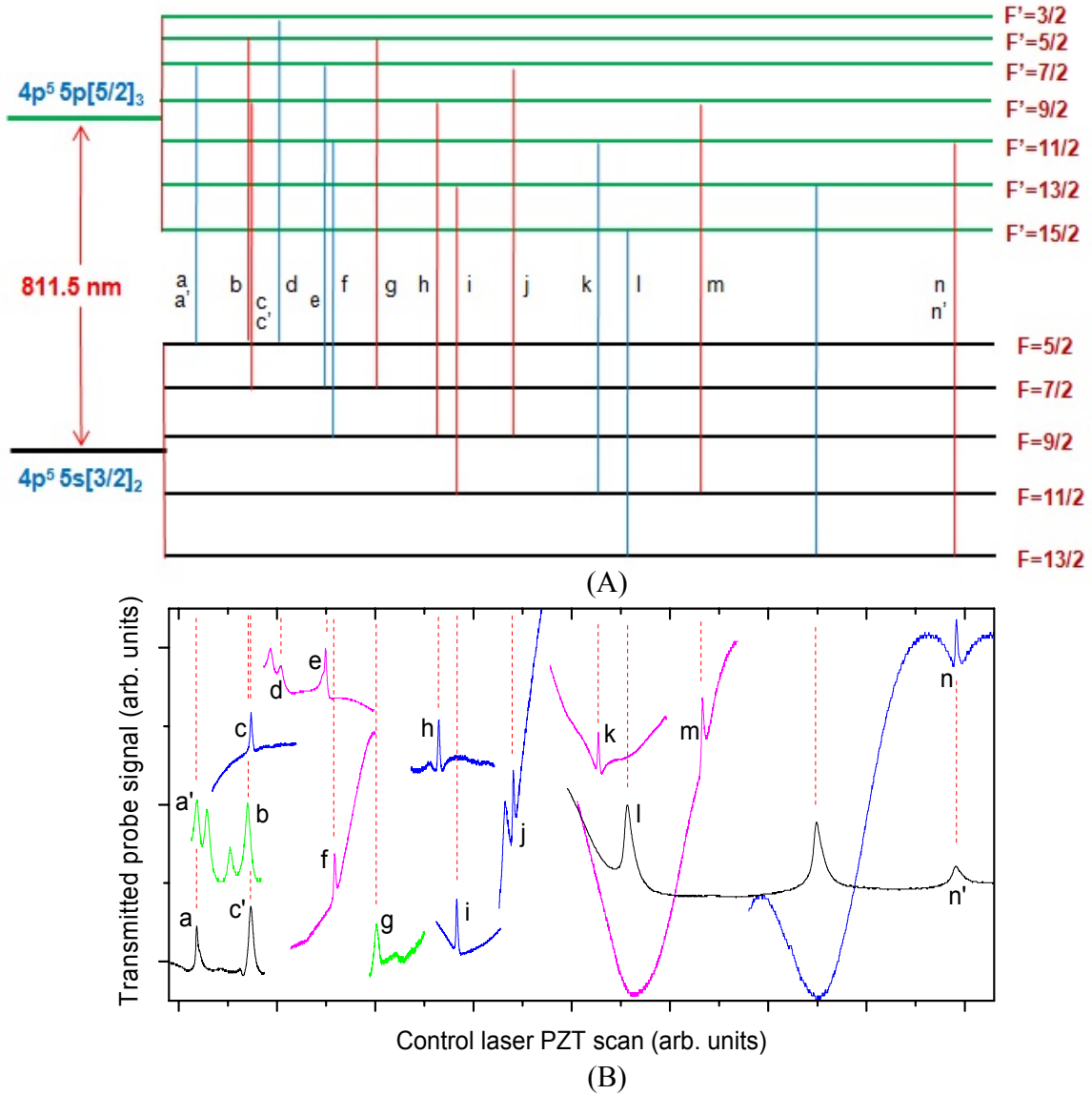

*Fig. 5: (A) The hyperfine levels in $4p^55s[3/2]_2$ and $4p^55p[5/2]_3$ manifolds of metastable $^{83}Kr$ atom. (B) Experimentally observed EIT signals when control laser frequency is scanned and probe laser is kept at a resolved transition in SAS-p. Here a sharp peak in the transmitted probe signal (i.e. EIT peak) indicates the control laser being resonant to a hyperfine transition making Λ- or V-type system with the probe laser hyperfine transition. The frequency corresponding to these EIT peaks in 5 (B) is mentioned in Table III. The red colour lines in Fig. 5(A) indicate the transitions which are unresolved by SAS but resolved by EIT method.*

photodiode signal when probe laser is kept exactly at transition 7/2 - 7/2 using SAS-p. As expected, the Λ-type EIT has narrower line width ($\Gamma_{EIT}$ ~ 8 MHz) than that of V-type EIT ($\Gamma_{EIT}$ ~ 27 MHz). This is mainly because of smaller dephasing rate and lesser optical pumping in former case [32]. Similarly, other peaks are also obtained either employing Λ-type EIT configuration or V-type EIT configuration by keeping the probe laser frequency resonant to one of these transitions obtained in SAS-p. Thus, we have visibly resolved all the 15 transitions in $4p^55s[3/2]_2$ to $4p^55p[5/2]_3$ manifolds of $^{83}$Kr* using the EIT technique and results are evident in Fig.5 (B). The Table II (column 4) summarizes the frequency positions of various hyperfine transitions measured using our EIT method. These spectral positions of peaks are estimated after taking the average of frequency values obtained in several repeated measurements. The values are rounded-off to 1 MHz precision. Here column 1 (Table II) shows the transitions identified from the experimentally observed EIT signals. The control laser is scanned around the transition shown in column 1 and the probe laser is fixed at another transition shown in column 2 to make either Λ-type or V-type system in our method. The frequency positions of these hyperfine transitions (column 1) reported by other methods are also listed in column 5 and 6 of the table for the comparison purpose. The linewidths of our observed EIT signals (Λ-type) are 6 to 9 times narrower than the corresponding spectral width (~65 MHz) obtained by J. R. Brandenberger using a typical SAS method [31]. The linewidths of these EIT signals are also smaller than that reported by B. D. Cannon et al using ionization mass spectrometry [33]. We note here that our work is entirely based on the resonances originated by quantum interference and not by classical methods used in the previous works [31, 33]. In Λ-type EIT signals, the linewidths are not limited by the barrier of natural linewidth and sub-natural lineiwdth EIT signals are already reported earlier [19, 27, 35]. In our setup, linewidth of EIT signal can be further reduced by reducing the control beam power.

**Table II:** *The investigated hyperfine transitions in $4p^55s[3/2]_2$ to $4p^55p[5/2]_3$ manifolds of $^{83}$Kr* atom and their spectral positions with respect to $^{84}$Kr* transition frequency.*

| Investigated hyperfine transition (F - F') | Transition in resonance with probe laser (F - F') | Parameters of the investigated transition using EIT in this work | | Averaged Frequency position by SAS (MHz) | Frequency position as estimated in Ref [24] (MHz) |
| --- | --- | --- | --- | --- | --- |
| | | Position of EIT Peak in Fig.5 (B) | Averaged Frequency position by EIT (MHz) | | |
| 5/2 - 7/2 | 7/2 - 7/2 (Λ)<br>5/2 - 3/2 (V) | a<br>a' | *-1407* | -1407 | -1408 |
| 5/2 - 5/2 | 5/2 - 3/2 (V) | b | *-1147* | - | -1147 |
| 7/2 - 9/2 | 11/2 - 9/2 (Λ)<br>7/2 - 7/2 (V) | c<br>c' | *-1131* | | -1132 |
| 5/2 - 3/2 | 5/2 - 7/2 (V) | d | -979 | -979 | -979 |
| 7/2 - 7/2 | 5/2 - 7/2 (Λ) | e | -752 | -752 | -753 |
| 9/2 - 11/2 | 11/2 - 11/2 (Λ) | f | -706 | -706 | -706 |
| 7/2 - 5/2 | 7/2 - 7/2 (V) | g | *-494* | - | -492 |
| 9/2 - 9/2 | 7/2 - 9/2 (Λ) | h | *-175* | - | -176 |
| 11/2 - 13/2 | 13/2 - 13/2 (Λ) | i | -86 | - | -87 |
| 9/2 - 7/2 | 7/2 - 7/2 (Λ) | j | *204* | - | 203 |
| 11/2 - 11/2 | 9/2 - 11/2 (Λ) | k | 635 | 635 | 635 |
| 13/2 - 15/2 | 13/2 - 13/2 (V) | l | *783* | 783 | 783 |
| 11/2 - 9/2 | 7/2 - 9/2 (Λ) | m | *1167* | - | 1166 |
| 13/2 - 13/2 | 13/2 - 13/2 | - | *1745* | 1745 | 1744 |
| 13/2 - 11/2 | 9/2 - 11/2 (Λ)<br>13/2 - 13/2 (V) | n<br>n' | 2466 | - | 2466 |

The frequency positions of various hyperfine resonances in $^{83}$Kr* resolved in this work with respect to resonant transition frequency of $^{84}$Kr* are listed in Table II along with the corresponding values reported in the previous work [24]. The frequency of a hyperfine level with respect to the ground state is given as

$$\nu_F = \nu_J + A\frac{C}{2} + B\frac{\frac{3}{4}C(C+1) - I(I+1)J(J+1)}{2I(2I-1)J(2J-1)} \qquad (3)$$

where $\nu_J$ is the frequency of unperturbed fine structure level for the known value of $J$, $I = 9/2$ for $^{83}$Kr and $C = F(F+1)-J(J+1)-I(I+1)$. For metastable $^{83}$Kr, the values of J are 2 and 3 for lower (4p$^5$5s[3/2]$_2$) and upper (4p$^5$5p[5/2]$_3$) manifolds respectively. The hyperfine number $F$ ranges from 5/2 to 13/2 for the lower manifold and 3/2 to 15/2 for the upper manifold. In Eq. (3), the parameters A and B are magnetic hyperfine constant and electric quadrupole hyperfine constant for a manifold with given J value.

The values of A and B for the lower and upper manifolds can be calculated in several ways if hyperfine transition frequencies between two manifolds are known [31, 33, 37]. In the present work we have used the linear transformation method as implemented by Parker et al.

earlier [37]. This method gives the best values of A and B, when evaluation of A and B is over determined due to excess of linear equations relating A and B. The linear equations for A and B are obtained as following by knowing the hyperfine splitting in adjacent levels in a manifold and using Eq. (3),

$$\Delta \nu_i = \alpha_i A + \beta_i B = y_i \qquad (4)$$

where $\Delta \nu_i = \nu_{F+1} - \nu_F = y_i$ is the hyperfine splitting between two adjacent energy levels in a manifold and index i varies from 1 to N-1 for N hyperfine levels in the manifold. The $y_i$ can be obtained by taking the difference of frequencies between two EIT peaks corresponding to two different hyperfine transitions involving a common hyperfine state. The values of $\alpha_i$ and $\beta_i$ depend on the value of F. Now defining the matrix vectors as

$$\alpha = \begin{pmatrix} \alpha_1 \\ \alpha_2 \\ . \\ . \\ \alpha_{N-1} \end{pmatrix} \text{ and } \beta = \begin{pmatrix} \beta_1 \\ \beta_2 \\ . \\ . \\ \beta_{N-1} \end{pmatrix}, \qquad (5)$$

the Eq. (4) can be rewritten in matrix form as

$$X_{(N-1)\times 2} \begin{pmatrix} A \\ B \end{pmatrix}_{2\times 1} = Y_{(N-1)\times 1}, \qquad (6)$$

where $X_{(N-1)\times 2} = (\alpha, \beta)$. To find A and B from Eq. (6), Moore-Penrose inversion is given as

$$\begin{pmatrix} A \\ B \end{pmatrix}_{2\times 1} = (X^T X)^{-1} X^T Y. \qquad (7)$$

In the present work, we have estimated the parameters A and B for the manifolds of the $^{83}$Kr* by using Eq. (7) and knowing the hyperfine splitting (Y) in the manifold from the experimentally measured frequency positions of EIT signals. To obtain hyperfine splitting accurately, the measurement of the EIT peak positions were repeated for several times and the average values are used. Table III shows the values of A and B as we obtained in this work, along with those obtained in the previous theoretical [38] and experimental work [31, 33, 34]. Our values of A and B obtained using EIT method agrees well with previously reported values in Ref. [34]. For 4p$^5$5s[3/2]$_2$ manifold, our (A, B) values differ from their values by (~ 0.02 %, ~ 0.11 % ), whereas for 4p$^5$5p[5/2]$_3$ manifold, our (A, B) values differ from their values by (~ 0.08 %, ~ 0.05 % ).

The uncertainties δA and δB (in A and B) have been evaluated when Y is replaced with δY in Eq. (7). We have constructed the vector δY by calculating the standard deviation in experimentally measured values of $y_i$ in the repeated measurements. Our estimated values of δA and δB are $5 \times 10^{-2}$ MHz and $1 \times 10^{-1}$ MHz respectively for $4p^55s[3/2]_2$ manifold and $4 \times 10^{-2}$ and $2 \times 10^{-1}$ respectively for $4p^55p[5/2]_3$ manifold (see numbers in parentheses in Table-III). These uncertainty values are much smaller than the values reported in the previous work in Refs. [31] and [33], and are comparable to the values reported in Ref. [34].

*Table III:* Measured magnetic hyperfine constants (A) and the electric quadrupole hyperfine constants (B) for $4p^55s[3/2]_2$ manifold and $4p^55p[5/2]_3$ manifold of $^{83}$Kr*. The number shown in the parentheses after each value represents the uncertainty in the last digit(s) of the respective value.

| State | Coefficients (in MHz) | Estimated by experiment | | | | Theory |
|---|---|---|---|---|---|---|
| | | This work | Ref. [34] | Ref. [33] | Ref. [31] | Ref. [38] |
| $4p^55s[3/2]_2$ | A | -243.92(5) | -243.87(5) | -243.93(4) | | -239.54 |
| | B | -453.5(1) | -453.1(7) | -452.93(60) | | -449.69 |
| $4p^55p[5/2]_3$ | A | -103.81(4) | -103.73(7) | -104.02(6) | -103(1) | -103.13 |
| | B | -439.0(2) | -438.8(12) | -436.9(17) | -430(30) | -431.70 |

## V. CONCLUSION

We have observed narrow EIT resonances in metastable $^{83}$Kr ($^{83}$Kr*) atom. The EIT resonances have been used to resolve the hyperfine transitions in $^{83}$Kr* which were not clearly resolved in SAS based spectroscopy work. The precise frequency position of EIT signals and their narrow linewidths has been exploited for the accurate measurement of hyperfine transition frequency. Using these results, the magnetic hyperfine constant (A) and the electric quadrupole hyperfine constant (B) for the manifolds $4p^55s[3/2]_2$ and $4p^55p[5/2]_3$ of $^{83}$Kr* atom are determined. The values of A and B parameters obtained in this work have been compared with those reported in the previous works. Our EIT based method can be useful in the resolution of spectrally rich transitions in other atoms and molecules also, including the hyperfine transitions in other noble gas atoms [39, 40] and weak transitions in heavy atoms [41].

*Acknowledgments*
We are thankful to Sethuraj K. R. for his help in the experiments. YBK is also thankful to Department of Atomic Energy for providing the post doctoral fellowship.

**References:**

[1] K. J. Boller, A. Imamoglu, and S. E. Harris, Phys. Rev. Lett. **66**, 2593 (1991).

[2] M. Fleischhauer, A. Imamoglu, and J. P. Marangos, Rev. Mod. Phys. **77**, 633 (2005).

[3] D. F. Phillips et al., Phys. Rev. Lett. **86**, 783 (2001).

[4] V. Wong, R. W. Boyd, C. R. Stroud Jr., R. S. Bennink, and A. M. Marino, Phys. Rev. A **68**, 012502 (2003).

[5] L.V. Hau, S. E. Harris, Z. Dutton, and C. H. Behroozi, Nature (London) **397**, 594 (1999).

[6] M. Mitsunaga, M. Yamashita, and H. Inoue, Phys. Rev. A **62**, 013817 (2000).

[7] S. Shahidani, M. H. Naderi, and M. Soltanolkotabi, Phys. Rev. A **88**, 053813 (2013).

[8] Y. Tamayama, K. Yasui, T. Nakanishi, and M. Kitano, Phys. Rev. B **89**, 075120 (2014).

[9] D. Petrosyan, J. Otterbach and M. Fleischhauer, Phys. Rev. Lett. **107**, 213601 (2011).

[10] J. Ghosh, R. Ghosh, F. Goldfarb, J.-L. Le Gouët, and F. Bretenaker, Phys. Rev. A **80**, 023817 (2009).

[11] B. Lubotzky, D. Shwa, T. Kong, N. Katz, and G. Ron, arXives: 1402.6275 (2014).

[12] Y. Liu, M. Davanc, V. Aksyuk, and K. Srinivasan, Phys. Rev. Lett. **110**, 223603 (2013).

[13] W. Li, D. Viscor, S. Hofferberth, and I. Lesanovsky, Phys. Rev. Lett. **112**, 243601 (2014).

[14] Y. Lin, J. P. Gaebler, T. R. Tan, R. Bowler, J.D. Jost, D. Leibfried, and D. J. Wineland, Phys. Rev. Lett. **110**, 153002 (2013).

[15] M. Klein, M. Hohensee, D. F. Phillips, and R. L. Walsworth, Phys. Rev. A **83**, 013826 (2011).

[16] U. Schnorrberger, J. D. Thompson, S. Trotzky, R. Pugatch, N. Davidson, S. Kuhr, and I. Bloch, Phys. Rev. Lett. **103**, 033003(2009).

[17] G. Heinze, C. Hubrich, and T. Halfmann, Phys. Rev. Lett. **111**, 033601(2013).

[18] V. I. Yudin, A. V. Taichenachev, Y. O. Dudin, V. L. Velichansky, A. S. Zibrov, and S. A. Zibrov, Phys. Rev. A **82**, 033807 (2010).

[19] Y. B. Kale, A. Ray, R. D'Souza, Q. V. Lawande, and B. N. Jagatap, App. Phys. B **100**, 505 (2010).

[20] M. Jain, H. Xia, G. Y. Yin, A. J. Merriam, and S. E. Harris, Phys. Rev. Lett. **77**, 4326 (1996).

[21] A. I. Lvovsky, B. C. Sanders, and W. Tittel, Nat. Photon. **3**, 706 (2009).

[22] C. Dorman and J. P. Marangos, Phys. Rev. A. **58**, 4121 (1998).

[23] C. Y. Chen, Y. M. Li, K. Bailey, T. P. O'Connor, L. Young, and Z.-T. Lu, Science **286**, 1139 (1999).

[24] A. I. Ludin, and B. E. Lehmann, App. Phys. B **61**, 461 (1995).

[25] K. S. Krane, "Introductory Nuclear Physics", Wiley (1987).

[26] A. Javan, O. Kocharovskaya, H. Lee, and M. O. Scully, Phys. Rev. A **66**, 013805 (2002).

[27] Y. B. Kale, A. Ray, N. Singh, Q. V. Lawande, and B. N. Jagatap, E. Phys. J D **61**, 221 (2010).

[28] C. Goren, A. D. Wilson-Gordon, M. Rosenbluh, and H. Friedmann, Phys. Rev. A **67**, 033807 (2003).

[29] http://steck.us/alkalidata/rubidium85numbers.pdf

[30] http://www.kip.uni-heidelberg.de/matterwaveoptics/teaching/practical/F65_EIT_Anleitung.pdf

[31] J. R. Brandenberger Phys. Rev. A **39,** 64 (1989).

[32] D. J. Fulton, S. Shepherd, R. R. Moseley, B. D. Sinclair, and M. H. Dunn, Phys. Rev. A Phys. Rev. A **52**, 2302 (1995).

[33] B. D. Cannon and G. R. Janik, Phys. Rev. A **42**. 397 (1990).

[34] B. D. Cannon, Phys. Rev. A. **47**, 1148 (1993).

[35] S. M. Iftiquar, G. R. Karve, and V. Natarajan, Phys. Rev. A **77**, 063807 (2008).

[36] T. Trickl, M. J. J. Vrakking, E. Cromwell, Y. T. Lee, and A. H. Kung, Phys. Rev. A **39**, 3048 (1989).

[37] S. C. Parker and J. R. Brandenberger, Phys. Rev. A **44**, 3354 (1991).

[38] X. Husson and J.-P. Grandin, J. Phys. B **12**, 3883 (1979).

[39] J. Welte, I. Steinke, M. Henrich, F. Ritterbusch, M. K. Oberthaler, W. Aeschbach-Hertig, W. H. Schwarz, and M. Trieloff, Rev. Sci. Instrum. **80**, 113109 (2009).

[40] E. Pawelec, S. Mazouffre and N. Sadeghi, Spectrochimica Acta Part B **66**, 470 (2011).

[41] D. Kilbane and G. O'Sullivan, Phys. Rev. A **82**, 062504 (2011).